\documentclass[10pt,a4paper,twoside]{article}
\usepackage{epsfig}
\usepackage{baltlat6}
\usepackage{array}
\usepackage{here}
\usepackage{natbib}
\usepackage{hyperref}

\makeatletter
\def\captionb#1#2{\refstepcounter{figure}\small\vspace{2mm}
\vbox{\baselineskip=9.5pt\hhuad{\bfseries Fig.\,#1.}
 \,#2} \baselineskip=11pt}

\def\sectionb#1#2{\refstepcounter{section}\vspace{5mm}\hbox{\kern-.9pt
{#1.\ \ }\vtop{\noindent #2}\nopagebreak}
\vspace{1mm} \noindent\baselineskip=11pt}

\def\subsectionb#1#2{\refstepcounter{subsection}\vspace{3mm}\hbox{\kern-.9pt
{\it #1.\ \ }\vtop{\noindent\it #2}\nopagebreak}
\vspace{.5mm} \noindent\baselineskip=11pt}

\def\subsubsectionb#1#2{\refstepcounter{subsubsection}\vspace{3mm}\hbox{
\kern-.9pt
{#1.\ \ }\vtop{\noindent #2}\nopagebreak}
\vspace{.5mm} \noindent\baselineskip=11pt}
\makeatother

\bibpunct{(}{)}{;}{a}{}{,}
\begin{document}
\ \
\vspace{0.5mm}
\setcounter{page}{119}
\vspace{8mm}

\titlehead{Baltic Astronomy, vol.\,25, 119--138, 2016}

\titleb{REVEALING THE NETWORK OF PERIODIC ORBITS IN\\ GALAXY MODELS
WITH A PROLATE OR AN OBLATE DARK\\ MATTER HALO COMPONENT}

\begin{authorl}
\authorb{Euaggelos E. Zotos}{}
\end{authorl}

\moveright-3.2mm
\vbox{
\begin{addressl}
\addressb{}{Department of Physics,  School of Science, Aristotle
University of Thessaloniki,\\
GR-541 24, Thessaloniki, Greece;  e-mail: evzotos@physics.auth.gr}
\end{addressl}}

\submitb{Received: 2015 April 27; accepted: 2015 June 23}

\begin{summary} Locating the position of periodic orbits in galaxies is
undoubtedly an issue of paramount importance.  We reveal the
position and the stability of periodic orbits of stars moving in the
meridional plane $(R,z)$ of an axially symmetric galactic model with
a disk, a spherical nucleus, and a biaxial dark matter halo
component.  In particular, we study how all the involved parameters
of the dynamical system influence the position and the stability of
all resonant families.  To locate the position and measure the
stability of periodic orbits we use a highly sensitive numerical
code which is able to identify resonant periodic orbits of the type
$n:m$.  Two cases are studied for every parameter:  (i) the case
where the dark matter halo component is prolate and (ii) the case
where an oblate dark matter halo is present.  Our numerical
exploration reveals that all the dynamical quantities affect, more
or less, the position and the stability of the periodic orbits.  It
is shown that the mass of the nucleus, the mass of the disk, the
halo flattening parameter, the scale length of the halo, the angular
momentum, and the total orbital energy are the most influential
quantities, while the effect of all other parameters is much weaker.
\end{summary}

\begin{keywords}
galaxies: kinematics and dynamics -- structure -- periodic orbits
\end{keywords}

\resthead{Periodic orbits in galaxies with dark matter haloes}
{Euaggelos E. Zotos}

\sectionb{1}{INTRODUCTION}
\label{intro}
\vskip-2mm

Periodic orbits undoubtedly play an important role in the analysis
of various types of dynamical systems.  Indeed, in dynamical systems
with strong chaotic behavior, unstable periodic orbits create a
``skeleton" for chaotic trajectories \citep{C91}. Furthermore, a
well regarded definition of chaos \citep{D89} requires the existence
of an infinite number of unstable periodic orbits that are dense in
the chaotic set.

Several geometric and dynamical properties of chaotic sets, such as
Lyapunov exponents, fractal dimensions, entropies \citep{O93}, can
be determined from the location and the stability properties of the
embedded periodic orbits.  Periodic orbits are central to
understanding of quantum-mechanical properties of non-separable
systems: the energy level density of such systems can be expressed
in a semiclassical approximation as a sum over the unstable periodic
orbits of the corresponding classical system \citep{G90}.
Topological description of a chaotic attractor also benefits from
the knowledge of periodic orbits.  For example, a large set of
periodic orbits is highly constraining to the symbolic dynamics and
can be used to extract the location of a generating partition
\citep{DLBD00,PL00}.  The significance of periodic orbits for the
experimental study of dynamical systems has been demonstrated in a
wide variety of systems \citep{LK89}, especially for the purpose of
controlling chaotic dynamics \citep{OGY90} with possible application
in communication \citep{BLG97}.  In a Hamiltonian system, periodic
orbits are not usually isolated but form one-parametric families.
Naturally, the value of the Hamiltonian function plays the role of
the parameter.  Thus even in the case when the original Hamiltonian
does not explicitly contain any parameter, it is possible to observe
bifurcations of periodic orbits.  A bifurcation corresponds to a
resonance between the frequency of a periodic orbit and the
frequency of small oscillations around it.  In a generic situation
there is a family of hyperbolic periodic orbits of a multiple
period, which shrinks to the resonant periodic orbit at an exact
resonance. Separatrices of the hyperbolic periodic orbit have to
intersect due to Hamiltonian nature of the problem.  Segments of
separatrices of the corresponding resonant normal form make up a
closed loop around the periodic trajectory.

In an effort to understand the structure of the solutions of
non-integrable dynamical systems, the numerical determination of
their periodic solutions and their stability properties plays a role
of fundamental importance.  The fact that for most dynamical systems
the periodic orbits are dense in the set of all possible solutions,
at least in certain parts of the phase space, necessitates the
presence of an efficient numerical method for their determination.
Over the last decades, several works developing different numerical
algorithms in order to compute periodic orbits have been presented.
\citet{A61,A63} in pioneer works examined in detail nearly circular
periodic orbits inside a heterogeneous ellipsoid for two particular
cases regarding the density distribution.  Furthermore, \citet{A74}
developed a topological method of finding periodic orbits.  Later
on, Prendergast proposed a method to find approximate solutions of
the Hamiltonian equations of motion in the form of rational
formulae. \citet{P82} and \citet{W84} applied this method to
Hamiltonian systems with one and two degrees of freedom, while
\citet{CS90} deployed the same method to the logarithmic potential
in order to obtain the families of periodic orbits.  The use of
differential correction algorithms for the numerical computation of
either two- or three-dimensional periodic orbits is not a new
result. The contributions of \citet{DH67} and \citet{HL86} with
respect to the systems with two degrees of freedom, and
\citet{BLO94}, Contopoulos \& Barbanis (1985), \citet{H84},
\citet{KS89}, \citet{LP02} and \citet{S99} with respect to the
systems with three degrees of freedom, can be mentioned. Usually, a
dynamical system admits some kind of symmetry, and, consequently,
the traditional approaches used for computing periodic orbits were
based on those symmetries. However, when dealing with force fields
without symmetries, a different approach must be used. The normal
procedure is then the application of the Poincar\'{e} map, and
differential corrections are obtained through the computation of the
state transition matrix along the periodic orbit. On the other hand,
for conservative systems the monodromy matrix has one unit
eigenvalue with multiplicity two related to the time invariance of
the system thus preventing the computation of the corrections.  Two
approaches are normally used to compute the nontrivial eigenvalues,
both of them based on the integration of the variations in Cartesian
coordinates. The first computes the complete state transition matrix
and uses basic techniques of matrix algebra for obtaining the
eigenvalues of a singular matrix.  The second eliminates the two
unit eigenvalues from the system by creating a lower dimensional
map.  In a recent paper by \citet{Z14} (hereafter Paper I) we
introduced a composite analytic, axially symmetric galactic
gravitational model that embraces the general features of a disk
galaxy with a dense massive nucleus and a biaxial prolate or oblate
dark matter halo component. Then we distinguished between regular
and chaotic motion of stars moving in the meridional $(R,z)$ plane
and we also performed an orbit classification by separating regular
orbits into different regular families.  In the present paper we use
the same galactic model in an attempt to reveal the complete network
of periodic orbits.

This paper is organized as follows.  In Section \ref{galmod}, we
present in detail the structure and the properties of the
gravitational galactic model.  In Section \ref{comd}, we describe
some theoretical and computational details of our methods.  In the
following section, we investigate how all the involved parameters of
the dynamical system influence the position and the stability of the
periodic orbits when a prolate or an oblate dark halo component is
present.  Our paper ends with Section \ref{disc}, where the
discussion and the conclusions of this research are presented.

\sectionb{2}{PRESENTATION OF THE GALACTIC MODEL}
\label{galmod}
\vskip-2mm

Let us briefly recall the galactic gravitational model which was
introduced in Paper I. The total potential $V(R,z)$ consists of three
components:  a spherical nucleus, a disk, and a biaxial dark matter halo
component.

For the description of the nucleus, we use the Plummer potential
(e.g., Binney \& Tremaine 2008):
\begin{equation}
V_{\rm n}(R,z) = \frac{- G M_{\rm n}}{\sqrt{R^2 + z^2 + c_{\rm n}^2}}.
\label{Vn}
\end{equation}
Here, $G$ is the gravitational constant, while $M_{\rm n}$ and
$c_{\rm n}$ are the mass and the scale length of the nucleus,
respectively.  At this point, we must clarify that we do not include
any relativistic effects, because the nucleus represents a bulge
rather than a black hole or any other compact object.

The galactic disk is represented by a generalization of the \citet{MN75}
potential \citep[see also][]{CI87}
\begin{equation}
V_{\rm d}(R,z) = \frac{- G M_{\rm d}}{\sqrt{b^2 + R^2 + \left(\alpha + \sqrt{h^2
+ z^2}\right)^2}},
\label{Vd}
\end{equation}
where $M_{\rm d}$ is the mass of the disk, $b$ is the core radius of the
disk-halo, $\alpha$ is the scale length of the disk, while $h$
corresponds to the disk scale height.

The potential of the dark matter halo is modeled by the flattened
axisymmetric logarithmic potential
\begin{equation}
V_{\rm h}(R,z) = \frac{\upsilon_0^2}{2}\ln \left(R^2 + \beta z^2 + c_{\rm h}^2
\right),
\label{Vh}
\end{equation}
where $\beta$ is the flattening parameter and $c_{\rm h}$ stands for
the scale length of the dark halo component.  The parameter
$\upsilon_0$ is used for the consistency of the galactic units. This
potential can model a wide variety of shapes of galactic haloes by
suitably choosing the parameter $\beta$.  In particular, when $0.1
\leq \beta < 1$, the dark matter halo is prolate, when $\beta = 1$
it is spherical and when $1 < \beta < 2$ it is oblate.  In this
work, we use a system of galactic units where the unit of length is
1 kpc, the unit of velocity is 10 km s$^{-1}$, and $G = 1$.  Thus,
the unit of mass is $2.325 \times 10^7 {\rm M}_\odot$, that of time
is $0.9778 \times 10^8$ yr, the unit of angular momentum (per unit
mass) is 10 km\,kpc$^{-1}$\,s$^{-1}$, and the unit of energy (per
unit mass) is 100 km$^2$\,s$^{-2}$.  Our main objective is to
investigate the regular or chaotic nature of orbits in two different
cases:  when the dark matter halo component is (i) prolate (PH
model) and (ii) oblate (OH model).  Our models have the following
standard values of the parameters:  $M_{\rm d} = 7000$
(corresponding to $1.63\times 10^{11}$ M$_\odot$, i.e., a normal
disk galaxy mass), $b = 6$, $\alpha = 3$, $h = 0.2$, $M_{\rm n} =
250$ (corresponding to $5.8\times 10^{9}$M$_\odot$), $c_{\rm n} =
0.25$, $\upsilon_0 = 20$ and $c_{\rm h} = 8.5$, while $\beta = 0.5$
for the PH model and $\beta = 1.5$ for the OH model.  The values for
the disk and the nucleus were chosen with a Milky Way-type galaxy in
mind \citep[e.g.,][]{AS91}.  In the case of the prolate dark halo,
the set of the values of the parameters defines the standard prolate
model (SPM), while when the dark halo is oblate, we use the standard
oblate model (SOM).  Here, we must point out that our gravitational
potential is truncated ar $R_{\rm max} = 50$ kpc, otherwise the
total mass of the galaxy modeled by this potential would be
infinite, which is obviously not physical.  The values of the
parameters of the standard models secure positive density inside the
region with $R \leq 50$ kpc\footnote{~The density of the model will
be negative at large enough distances from the center $(R > 50$
kpc), and this is a general property of all models with ellipsoidal
(but non-spherical) equipotentials.}.

Taking into account that the total potential $V(R,z)$ is axisymmetric,
the $z$-component of the angular momentum $(L_z)$ is conserved.  With
this restriction, orbits can be described by means of the effective
potential \citep[e.g.,][]{BT08}
\begin{equation}
V_{\rm eff}(R,z) = V(R,z) + \frac{L_z^2}{2R^2}.
\label{veff}
\end{equation}

The Hamiltonian to the effective potential given in Eq. (\ref{veff})
can be written as
\begin{equation}
H = \frac{1}{2} \left(\dot{R}^2 + \dot{z}^2 \right) + V_{\rm eff}(R,z) = E,
\label{ham}
\end{equation}
where $\dot{R}$ and $\dot{z}$ are momenta per unit mass, and conjugate
to $R$ and $z$, respectively, while $E$ is the numerical value of the
Hamiltonian, which is conserved.  Therefore, an orbit is restricted to
the permissible area in the meridional plane satisfying $E \geq V_{\rm
eff}$.

Consequently, the corresponding equations of motion in the
meridional plane are
\begin{eqnarray}
\ddot{R} = - \frac{\partial V_{\rm eff}}{\partial R}, \nonumber \\
\ddot{z} = - \frac{\partial V_{\rm eff}}{\partial z},
\label{eqmot}
\end{eqnarray}
where a double dot indicates a derivative with respect to time.

\sectionb{3}{THEORETICAL AND COMPUTATIONAL DETAILS}
\label{comd}
\vskip-2mm

For the location of the position and the measurement of the
stability of periodic orbits we use a highly sensitive numerical
code which is able to identify resonant periodic orbits of the type
$n:m$.  The $n:m$ notation we use for the resonant periodic orbits
is according to \citet{CA98} and \citet{ZC13}, where the ratio of
those integers corresponds to the ratio of the main frequencies of
the orbit, where main frequency is the frequency of greatest
amplitude in each coordinate.  Main amplitudes, when having a
rational ratio, define the resonances of an orbit.  Initially we
define for every variable parameter an interval of realistic values.
The numerical code begins from the value corresponding to the
standard model (SPM and SOM), which is always somewhere inside the
main interval, and then, using a variable step, scans all the
available interval thus calculating the entire resonant family.  In
order to check the robustness and efficiency of our numerical
algorithm, we recalculated some of the halo orbits of the classical
restricted three-body problem (R3BP) presented initially in
\citet{GLMS85}. All the performed tests indicated a satisfactory
agreement with the corresponding numerical outcomes for either
stability index of the period of the periodic orbits.

The equations of motion are solved for a given value of the Hamiltonian,
starting with initial conditions $(R_0,z_0,\dot{R_0},\dot{z_0})$ in the
plane $z = 0$, for $\dot{z} > 0$.  The next intersection with the $z =
0$ plane with $\dot{z} > 0$ is found and the exact initial conditions
for the periodic orbit are calculated using a Newton iterative method.
A periodic orbit is found when the initial and final coordinates
coincide with an accuracy of at least $10^{-12}$.

The estimation of the linear stability of a periodic orbit is based on
the theory of variational equations.  We first consider small deviations
from its initial conditions, and then integrate the orbit again to the
next upward intersection.  Let the variational equations of a specific
periodic orbit of period $T$ be
\begin{equation}
\dot{\xi_i}(t) = \displaystyle\sum_{j=1}^{4} \frac{\partial F_i}{\partial x_j}
\xi_j \ \ \ \ \ \ \ (i,j=1,2,3,4).
\label{varsys}
\end{equation}
If $X(t)$ is the matrix, whose columns are the four solutions of the
system (\ref{varsys}) with initial conditions (1,0,0,0), (0,1,0,0),
(0,0,1,0) and (0,0,0,1), then we have the so-called monodromy matrix
\begin{equation}
\displaystyle X(t) = \left(
\begin{array}{cccc}
\frac{\displaystyle \partial R}{\displaystyle \partial R_0} &
\frac{\displaystyle \partial R}{\displaystyle \partial z_0} &
\frac{\displaystyle \partial R}{\displaystyle \partial \dot{R_0}} &
\frac{\displaystyle \partial R}{\displaystyle \partial \dot{z_0
}} \\
\frac{\displaystyle \partial z}{\displaystyle \partial R_0} &
\frac{\displaystyle \partial z}{\displaystyle \partial z_0} &
\frac{\displaystyle \partial z}{\displaystyle \partial \dot{R_0}} &
\frac{\displaystyle \partial z}{\displaystyle \partial \dot{z_0
}} \\
\frac{\displaystyle \partial \dot{R}}{\displaystyle \partial R_0} &
\frac{\displaystyle \partial \dot{R}}{\displaystyle \partial z_0} &
\frac{\displaystyle \partial \dot{R}}{\displaystyle \partial \dot{R_0}}
& \frac{\displaystyle \partial \dot{R}}{\displaystyle \partial
\dot{z_0}} \\
\frac{\displaystyle \partial \dot{z}}{\displaystyle \partial R_0} &
\frac{\displaystyle \partial \dot{z}}{\displaystyle \partial z_0} &
\frac{\displaystyle \partial \dot{z}}{\displaystyle \partial \dot{R_0}}
& \frac{\displaystyle \partial \dot{z}}{\displaystyle \partial
\dot{z_0}}
\end{array}
\right).
\label{matrix}
\end{equation}

When $t = T$, there is also a monodromy matrix $X(T)$.  The
stability of a periodic orbit depends on the eigenvalues of this
monodromy matrix. We define the stability index (S.I.) as
\begin{equation}
\rm S.I. = Tr(X(T)) - 2, \label{SI}
\end{equation}
where $\rm Tr(X(T)) = \lambda_1 + \lambda_2 + \lambda_3 + \lambda_4$
is the trace of the monodromy matrix, while $\lambda_i$ $(i=1,4)$
are the eigenvalues.  Now, according to the value of S.I. we can
determine if a periodic orbit is stable or unstable.  In particular,
if $|\rm S.I.| < 2$ the periodic orbit is stable, if $|\rm S.I.| >
2$ the periodic orbit is unstable, while if $|\rm S.I.| = 2$ the
periodic orbit is critically stable.

For each initial condition of a periodic orbit, we integrated the
equations of motion (\ref{eqmot}), as well as the variational
equations (\ref{varsys}), using a double precision Bulirsch-Stoer
algorithm \citep[e.g.,][]{PTVF92} with a small time step of the
order of $10^{-2}$, which is sufficient enough for the desired
accuracy of our computations (i.e., our results practically do not
change by halving the time step).  Here we should emphasize, as our
previous numerical experience suggests, that the Bulirsch-Stoer
integrator is both faster and more accurate than a double precision
Runge-Kutta-Fehlberg algorithm of order 7 with the Cash-Karp
coefficients.  In all cases, the energy integral (Eq. \ref{ham}) was
conserved better than one part in $10^{-11}$, although for most
orbits it was better than one part in $10^{-12}$.

\sectionb{4}{THE NETWORK OF PERIODIC ORBITS}
\label{neto}
\vskip-2mm

In all cases, the energy was set to $600$ and the angular momentum
of the orbits is $L_z = 10$.  We chose, for both PH and OH models,
the particular energy level, which yields $R_{\rm max} \simeq 15$
kpc, where $R_{\rm max}$ is the maximum possible value of $R$ on the
$(R,\dot{R})$ phase plane.  In every case, we let only one quantity
vary in a predefined range, while fixing the values of all the other
parameters, according to SPM and SOM.  Once the values of the
parameters were chosen, the numerical code begins locating the
position and computing the corresponding stability index of periodic
orbits along the predefined range of values.  The step between two
successive values of a variable quantity is chosen in such a way
that every resonant family contains at least 5000 initial conditions
$(R_0,\dot{R_0})$ of periodic orbits with $z_0 = 0$, while
$\dot{z_0}$ is always obtained from the energy integral (Eq.
\ref{ham}).

Our numerical calculations indicate that in the PH models there are six
main resonant families:  (i) 2:1 banana-type orbits; (ii) 1:1 open
linear orbits; (iii) 4:3 resonant orbits; (iv) 6:3 resonant orbits; (v)
10:5 resonant orbits; (vi) 12:7 resonant orbits.  In addition, in the
OH models there are only five main resonant families:  (i) 2:1
banana-type orbits; (ii) 1:1 open linear orbits; (iii) 3:2 resonant
orbits; (iv) 4:3 resonant orbits; (v) 8:5 resonant orbits.  In Table 1
of Paper I we provide the types and the exact initial conditions of the
main orbital families for the SPM and SOM, while in Figs. 6 and 7 of
Paper I we depict the shapes of these resonant periodic orbits.

In every case we present a very informative diagram, the so-called
``characteristic" orbital diagram \citep{CM77}.  It shows the
evolution of the $R$ coordinate of the initial conditions of the
periodic orbits of each orbital family as a function of the value of
the variable quantity.  Here we should emphasize that, for periodic
orbits starting perpendicular to the $R$-axis, we need only the
initial condition of $R_0$ in order to locate them on the
characteristic diagram.  On the other hand, for orbits not starting
perpendicular to the $R$-axis, initial conditions as
position-velocity pairs $(R,\dot{R})$ are required and, therefore,
the characteristic diagram is now three-dimensional providing the
full information regarding the interrelations of the initial
conditions in a tree of families of periodic orbits.  In the
following two-dimensional characteristic diagrams we will see that
some resonant families cross one another.  Here it should be
clarified that the full characteristic curves of some resonant
families (i.e., 1:1, the 3:2 and 6:3) are three-dimensional since
$\dot{R_0} \neq 0$, however we decided to combine all families
together in a two-dimensional plot containing only the evolution of
the $R$ coordinate (of course, in the 3D $(R, \dot{R}, M_{\rm n })$
space the corresponding characteristic curves do not cross one
another).  Furthermore, we provide the so-called ``stability
diagram" \citep{CB85,CM85} which illustrates the stability of all
the families of periodic orbits in our dynamical system when the
numerical value of the variable quantity varies, while all the other
parameters remain constant.  A periodic orbit is stable if only the
stability index (S.I.)  \citep{Z13} is between --2 and +2.  This
diagram helps us monitor the evolution of S.I. of the resonant
periodic orbits as well as the transitions from stability to
instability and vice versa.

\begin{figure*}[!th]
\centerline{ \resizebox{0.90\hsize}{!}{\includegraphics{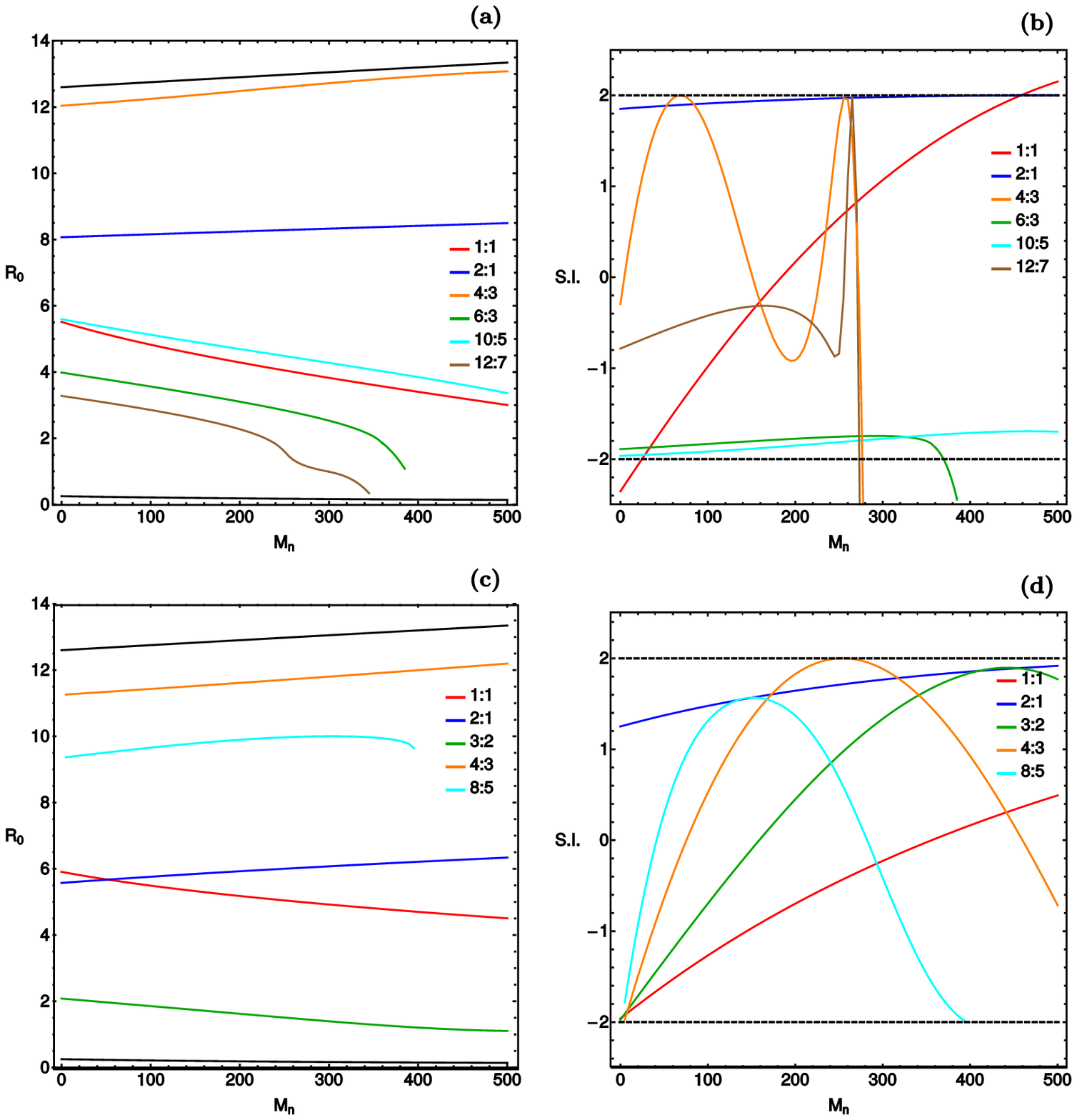}}}
\captionb{1}{Evolution of the $R$ coordinate of periodic orbits as a
function of mass of the nucleus, $M_{\rm n}$, for the PH models
(panel a) and OH models (panel c), and the evolution of the
stability index S.I. as a function of $M_{\rm n}$ for the PH models
(panel b) and OH models (panel d).} \label{fig1}
\end{figure*}

\subsectionb{4.1}{Influence of the mass of the nucleus}

To study how the mass of the nucleus, $M_{\rm n}$, influences the
position and stability of periodic orbits, we let it vary while
fixing all the other parameters of our model, and integrate orbits
in the meridional plane for the set $M_{\rm n} = \{0,50,
100,...,500\}$.  In all cases, the energy was set to $600$ and the
angular momentum of the orbits $L_z = 10$.  Once the values of the
parameters were chosen, we computed a set of initial conditions as
described in Sec.~\ref{comd}, and integrated the corresponding
orbits.

The evolution of the $R$ coordinate of the periodic orbits as
$M_{\rm n}$ varies for the PH models is shown in Fig.\,\ref{fig1}a.
The black solid lines indicate the borders of the limiting curves.
It can be seen that the main families of periodic orbits (i.e., the
2:1, 1:1, 4:3 and 10:5 families) are present throughout the entire
range of values of $M_{\rm n}$.  On the other hand, some secondary
resonances (i.e., the 6:3 and 12:7 families) terminate earlier than
the range of values.  In particular, our numerical calculations
shown in Fig.\,\ref{fig1}b reveal that the 6:3 family is unstable
for $371 \leq M_{\rm n} \leq 386$, while the 12:7 family is unstable
for $272 \leq M_{\rm n} \leq 344$.  Looking carefully at the
stability diagram of Fig.\,\ref{fig1}b we observe that the 1:1
family becomes unstable near the lower and upper limits of the
range.  Specifically the 1:1 resonant family is unstable for $0 \leq
M_{\rm n} \leq 23.8$ and for $457 \leq M_{\rm n} \leq 500$.  We also
see that the 2:1 resonant family spends most of its life near the
upper stability limit (+2), however it never crosses it.  Similarly
in Fig.\,\ref{fig1}c we present the evolution of the $R$ coordinate
of the periodic orbits as a function of $M_{\rm n}$ for the OH
models. Here the only resonant family, that terminates earlier at
relatively high values of the mass of the nucleus, is the 8:5
family.  In particular this family terminates at $M_{\rm n} = 395.2$
without exhibiting any instability first.

\begin{figure*}[!th]
\centerline{ \resizebox{0.90\hsize}{!}{\includegraphics{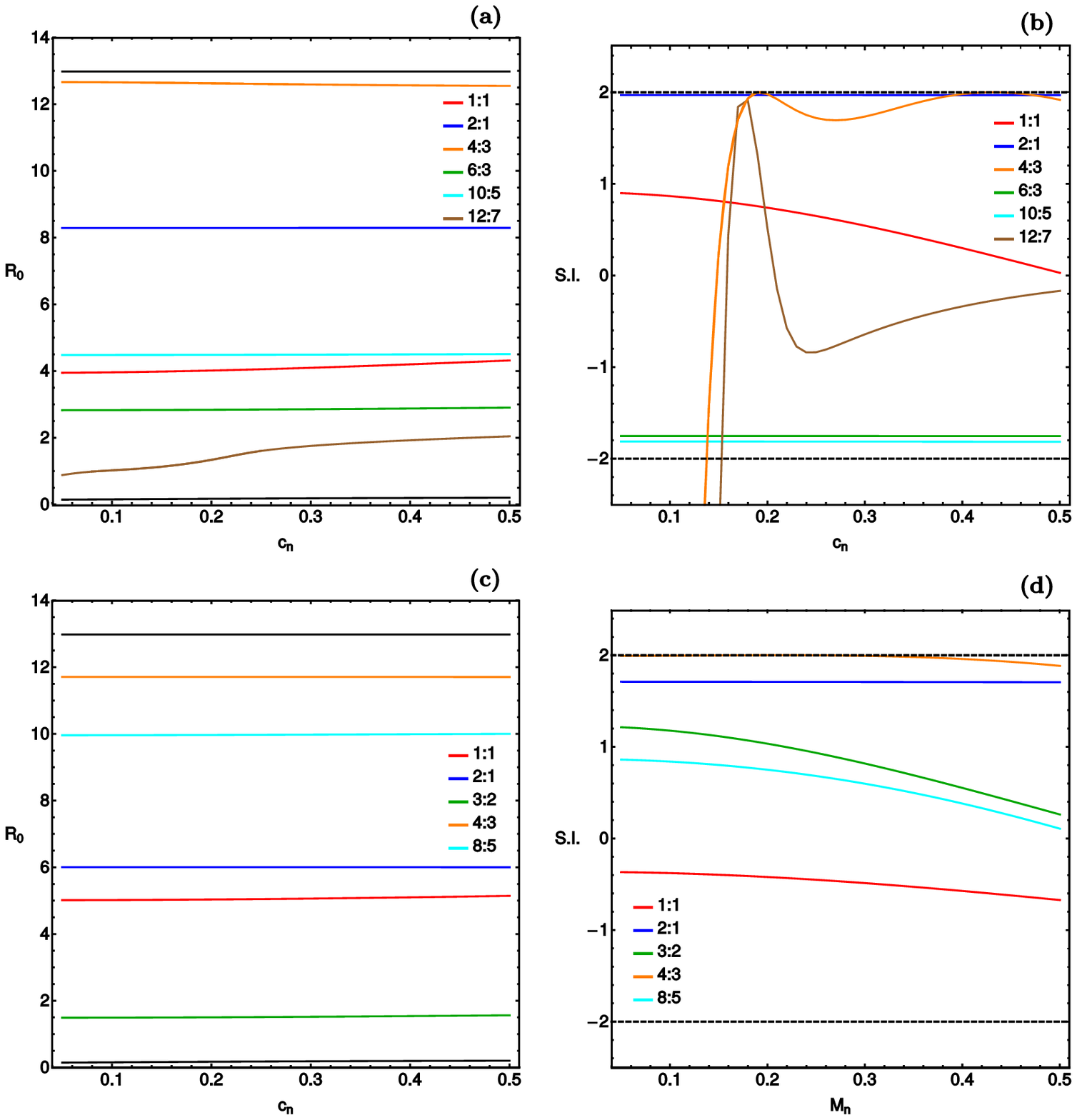}}}
\captionb{2}{Evolution of the $R$ coordinate of periodic orbits as a
function of the scale length of the nucleus, $c_{\rm n}$, for the PH
models (panel a) and OH models (panel c), and the evolution of the
stability index S.I. as a function of $c_{\rm n}$ for the PH models
(panel b) and OH models (panel d).} \label{fig2}
\end{figure*}

\subsectionb{4.2}{Influence of the scale length of the nucleus}

Now we proceed to investigate how the scale length of the nucleus
$c_{\rm n}$ influences the position and stability of the periodic orbits
in our PH and OH models.  Again, we let it vary while fixing all the
other parameters of our galactic model and integrating orbits in the
meridional plane for the set $c_{\rm n} = \{0.05,0.10,0.15,...,0.50\}$.

The evolution of the $R$ coordinate of the periodic orbits as
$c_{\rm n}$ varies for the PH models is shown in Fig.\,\ref{fig2}a.
It is evident that the scale length of the nucleus influences very
weakly the position of the periodic orbits in the PH models.  For
this reason all the resonant families survive the entire range of
values of $c_{\rm n}$.  The corresponding stability diagram shown in
Fig.\,\ref{fig2}b informs us that the 4:3 resonant family is
unstable for $0.05 \leq c_{\rm n} \leq 0.134$, while the 12:7
resonant family is unstable for $0.05 \leq c_{\rm n} \leq 0.147$.
Things are quite similar for the OH models. Indeed, the
characteristic diagram of Fig.\,\ref{fig2}c shows once more that the
scale length of the nucleus hardly affects the position of the
periodic orbits which, according to Fig.\,\ref{fig2}d, all remain
stable throughout the entire range of values of $c_{\rm n}$. It
should be noticed that the stability index of the 4:3 resonant
family evolves almost parallel to the upper stability limit (+2),
but never crosses it.

\begin{figure*}[!th]
\centerline{ \resizebox{0.90\hsize}{!}{\includegraphics{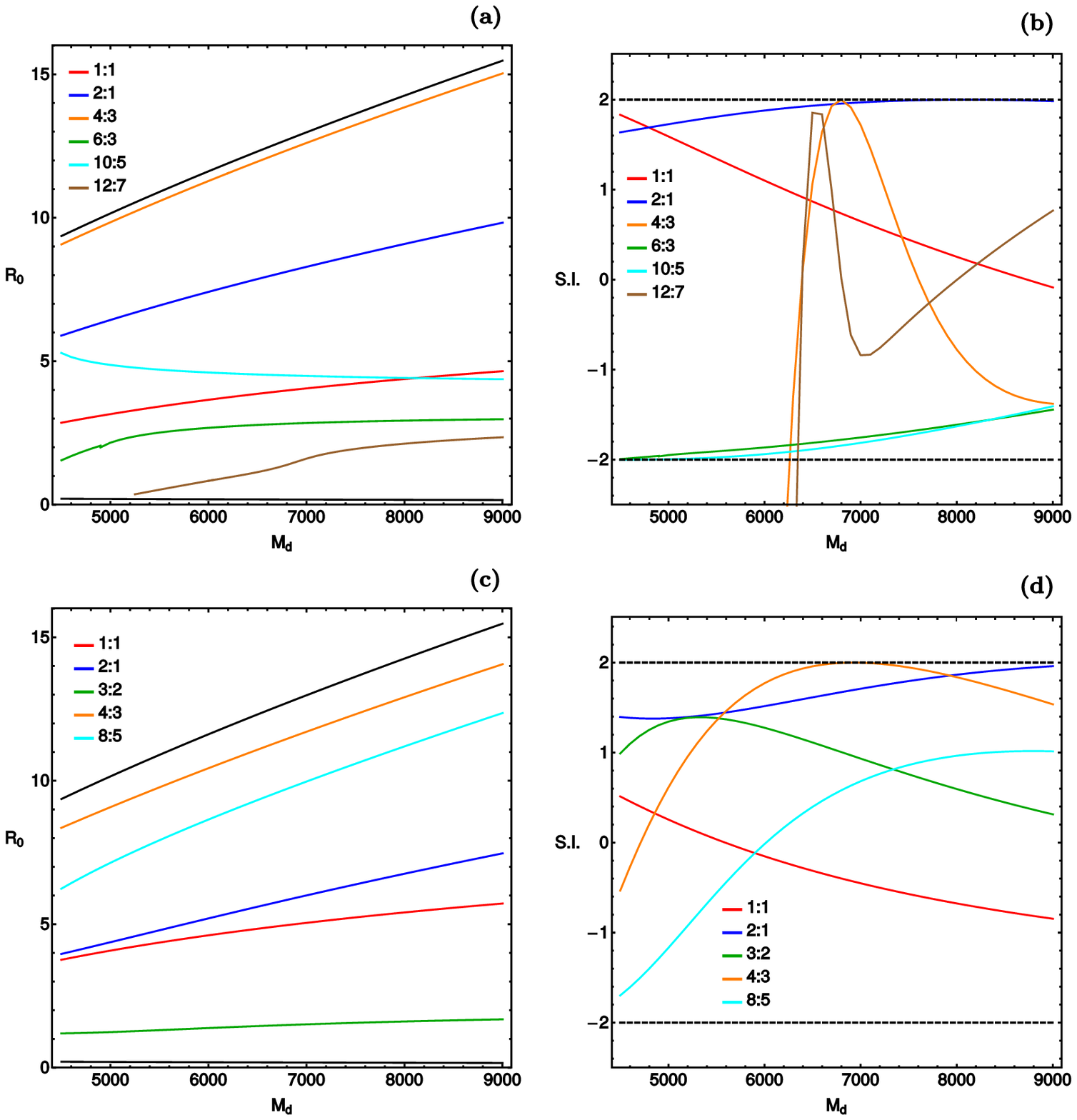}}}
\captionb{3}{Evolution of the $R$ coordinate of periodic orbits as a
function of the mass of the disk, $M_{\rm d}$, for the PH models
(panel a) and OH models (panel c), and the evolution of the
stability index S.I. as a function of $M_{\rm d}$ for the PH models
(panel b) and OH models (panel d).} \label{fig3}
\end{figure*}

\subsectionb{4.3}{Influence of the mass of the disk}

Our next step is to reveal how the position and stability of
periodic orbits in our PH and OH models are affected by the mass of
the disk $M_{\rm d}$.  As usual, we let this quantity vary while
fixing the values of all the other parameters of our galactic model
and integrating orbits in the meridional plane for the set $M_{\rm
d} = \{4500,5000,5500,...,9000\}$.

Fig.\,3a shows the evolution of the $R$ coordinate of periodic
orbits as a function of the mass of the disk for the PH models.  We
observe an almost linear relocation of the position of the periodic
orbits.  The corresponding stability diagram of Fig.\,\ref{fig3}b
reveals that the 4:3 resonant family is unstable for $4500 \leq
M_{\rm d} \leq 6218$, while the 12:7 resonant family is unstable for
$5248 \leq M_{\rm d} \leq 6295$ and terminates at $M_{\rm d} =
5248$.  Furthermore, we see that the 2:1 resonant family for about
$M_{\rm d} > 6800$ becomes almost critically stable.  The evolution
of the $R$ coordinate of the periodic orbits as a function of the
mass of the disk for the OH models is shown in Fig.\,\ref{fig3}c.
One can observe that all resonant families are present throughout
the entire range of the values of $M_{\rm d}$, while the
corresponding stability diagram of Fig.\,\ref{fig3}d indicates that
all resonant families are stable.

\begin{figure*}[!th]
\centerline{ \resizebox{0.90\hsize}{!}{\includegraphics{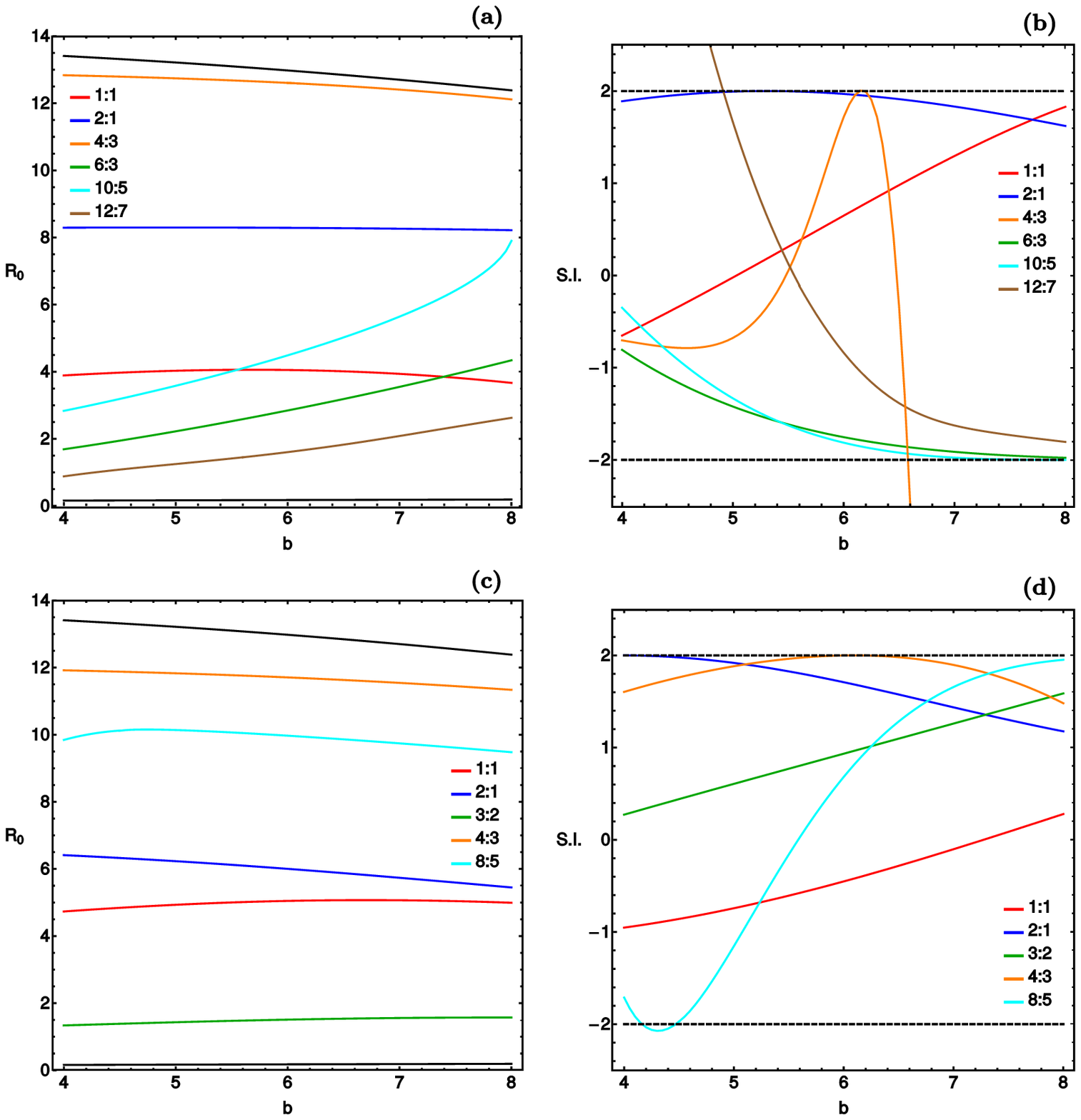}}}
\captionb{4}{Evolution of the $R$ coordinate of periodic orbits as a
function of the core radius of the disk-halo, $b$, for the PH models
(panel a) and OH models (panel c), and the evolution of the
stability index S.I. as a function of $b$ for the PH models (panel
b) and OH models (panel d).} \label{fig4}
\end{figure*}

\subsectionb{4.4}{Influence of the core radius of the disk-halo}

The next parameter under investigation is the core radius of the
disk-halo, $b$.  We will try to understand how the position and the
stability of the periodic orbits in our PH and OH galaxy models are
influenced by $b$.  Again, we let this quantity vary while fixing
the values of all the other parameters of our galactic model and
integrating orbits in the meridional plane for the set $b =
\{4,4.5,5,...,8\}$.

The evolution of the position of periodic orbits as a function of
the core radius of the disk-halo for the PH models is given in
Fig.\,\ref{fig4}a.  It is interesting to notice that for $b = 8$ the
10:5 secondary resonant family almost bifurcates from the main
parent 2:1 resonant family.  In addition, the corresponding
stability diagram shown in Fig.\,\ref{fig4}b indicates that the 4:3
resonant family becomes unstable for $6.58 \leq b \leq 8$, while the
12:7 resonant family is unstable for $4 \leq b \leq 4.94$.  In the
same vein we present in Fig.\,\ref{fig4}c the evolution of the $R$
coordinates of periodic orbits of the OH models as a function of
$b$.  Here the vast majority of the computed periodic orbits are
stable, while only the 8:5 resonant family displays a small branch
of unstable periodic orbits. Looking at Fig.\,\ref{fig4}d we see
that this branch is located near the lower boundary of the family
and specifically for $4.21 \leq b \leq 4.46$.

\begin{figure*}[!th]
\centerline{ \resizebox{0.90\hsize}{!}{\includegraphics{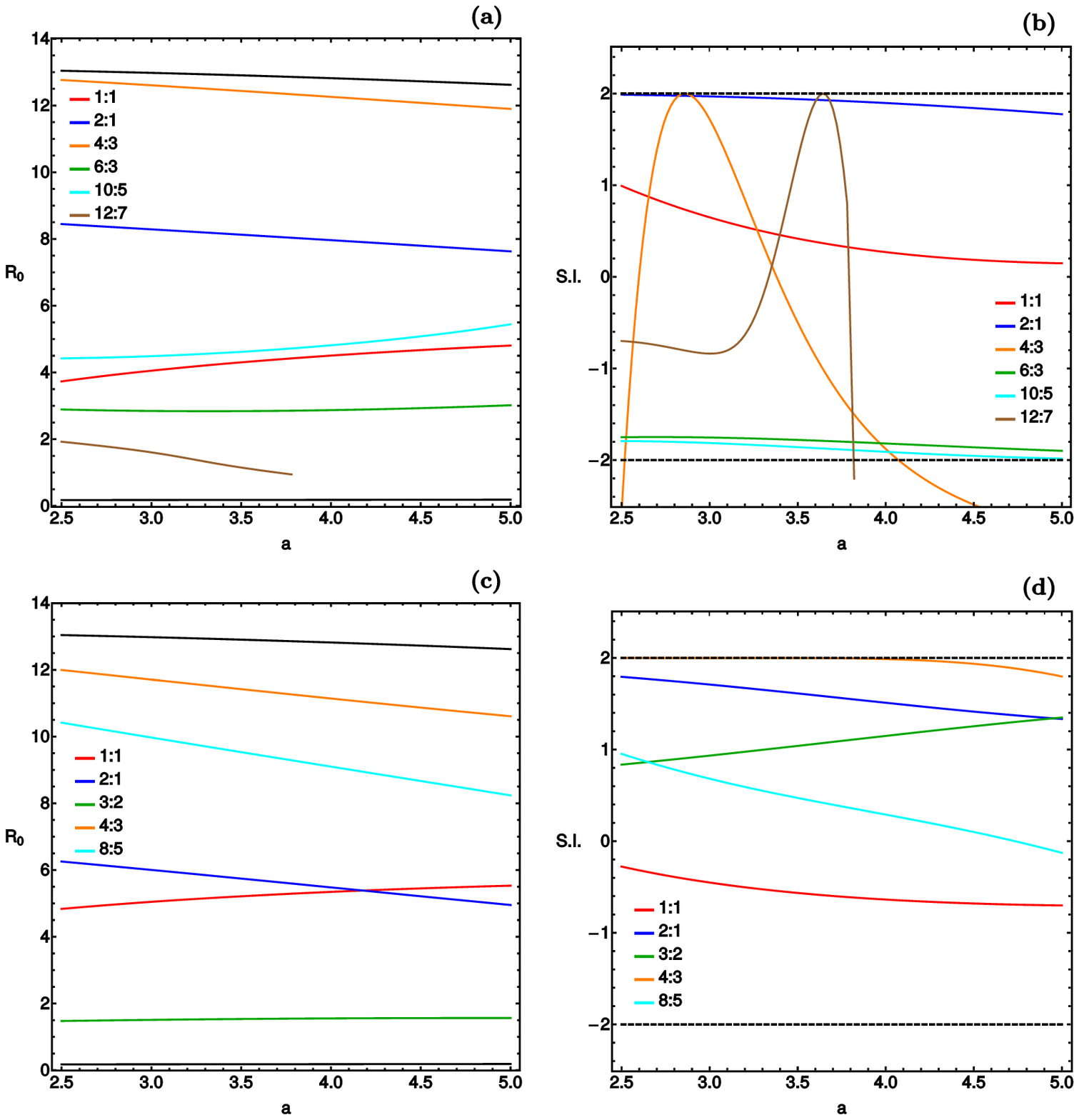}}}
\captionb{5}{Evolution of the $R$ coordinate of periodic orbits as a
function of the scale length of the disk, $\alpha$, for the PH
models (panel a) and OH models (panel c), and the evolution of the
stability index S.I. as a function of $\alpha$ for the PH models
(panel b) and OH models (panel d).} \label{fig5}
\end{figure*}

\subsectionb{4.5}{Influence of the scale length of the disk}

We continue our quest trying to understand how the scale length of a
galaxy, $\alpha$, influences  the position and the stability of
periodic orbits in our PH and OH galaxy models.  As usual, we let
this parameter vary while fixing the values of all the other
quantities of our galactic model and integrating orbits in the
meridional plane for the set $\alpha = \{2.5,2.75,3,...,5\}$.

The following Fig.\,\ref{fig5}a shows the evolution of the $R$
coordinate of periodic orbits as a function of the scale length of
the disk for the PH models.  We observe that the only resonant
family that terminates earlier is the 12:7 family.  In particular,
according to the corresponding stability diagram of
Fig.\,\ref{fig5}b this family first becomes unstable for $3.79 \leq
\alpha \leq 3.82$ and then it terminates.  Moreover, the 4:3
resonant family is also unstable in the intervals $2.5 \leq \alpha
\leq 2.52$ and $4.07 \leq \alpha \leq 5$. The evolution of the
position of all resonant families of periodic orbits as a function
of $\alpha$ for the OH models is given in Fig.\,\ref{fig5}c.  The
corresponding stability diagram of Fig.\,\ref{fig5}d suggests that
all resonant families are stable, while only the 4:3 family is
almost critically stable in the interval $2.5 \leq \alpha < 4.4$.

\begin{figure*}[!th]
\centerline{ \resizebox{0.90\hsize}{!}{\includegraphics{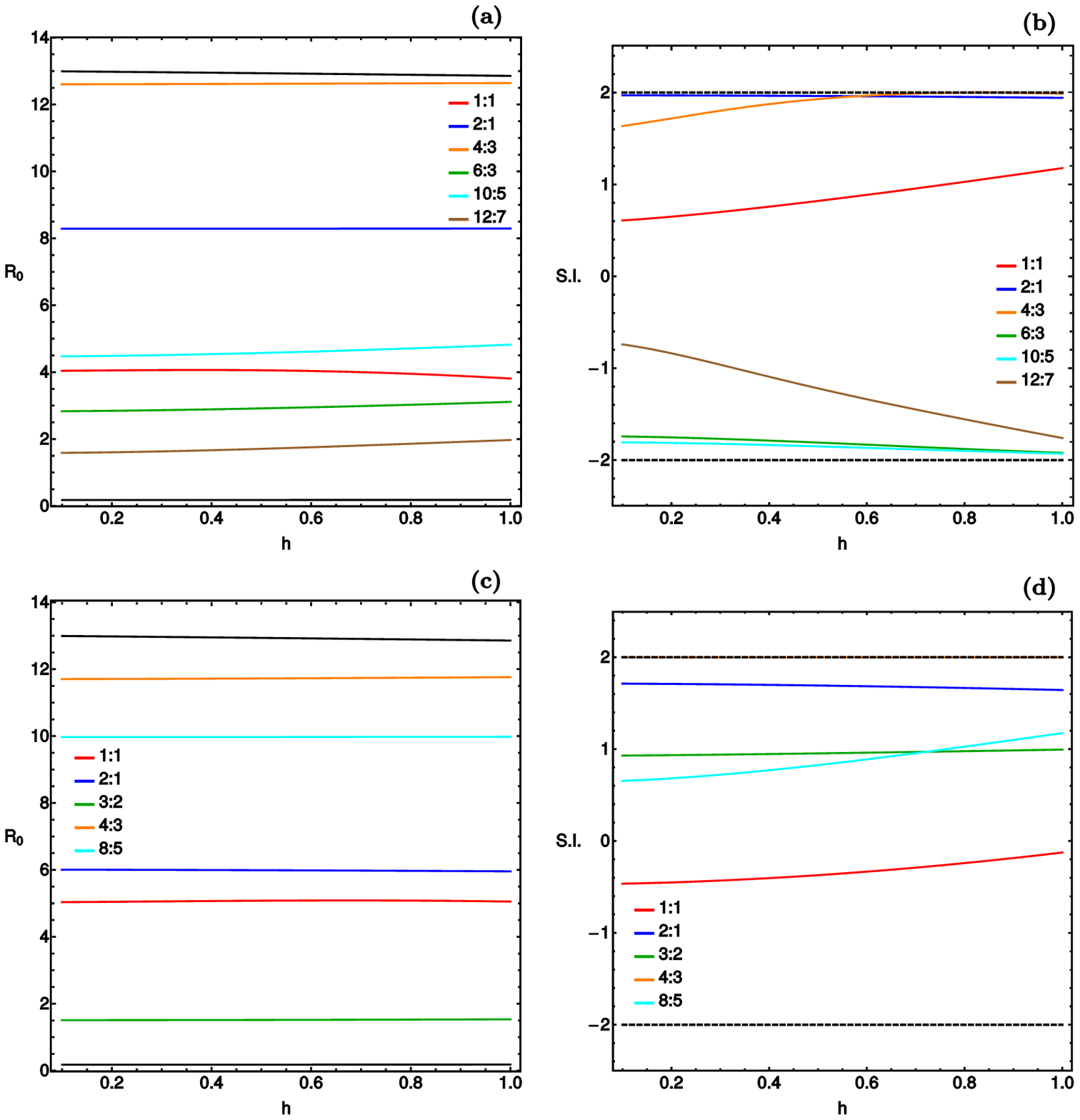}}}
\captionb{6}{Evolution of the $R$ coordinate of periodic orbits as a
function of the scale height of the disk, $h$, for the PH models
(panel a) and OH models (panel c), and the evolution of the
stability index S.I. as a function of $h$ for the PH models (panel
b) and OH models (panel d).} \label{fig6}
\end{figure*}

\subsectionb{4.6}{Influence of the scale height of the disk}

The next stop of our investigation is to determine how the position
and the stability of periodic orbits in our PH and OH galaxy models
are affected by the scale height of the disk, $h$.  Following the
usual procedure, we let this parameter vary while fixing the values
of all the other quantities of our galactic models and integrating
orbits in the meridional plane for the set $h =
\{0.1,0.2,0.3,...,1\}$.

The evolution of the position of periodic orbits as a function of
the scale height of the disk for the PH and OH models is given in
Figs.\,\ref{fig6}a and \ref{fig6}c, respectively.  We observe that
all the characteristic curves are monotone straight lines, implying
that the scale height of the disk practically does not influence the
position of the resonant periodic orbits.  The corresponding
stability diagrams shown in Figs.\,\ref{fig6}b and \ref{fig6}d
indicate that all resonant families are stable except for the 4:3
family which is almost critically stable in most parts of the
investigated interval.

\begin{figure*}[!th]
\centerline{ \resizebox{0.90\hsize}{!}{\includegraphics{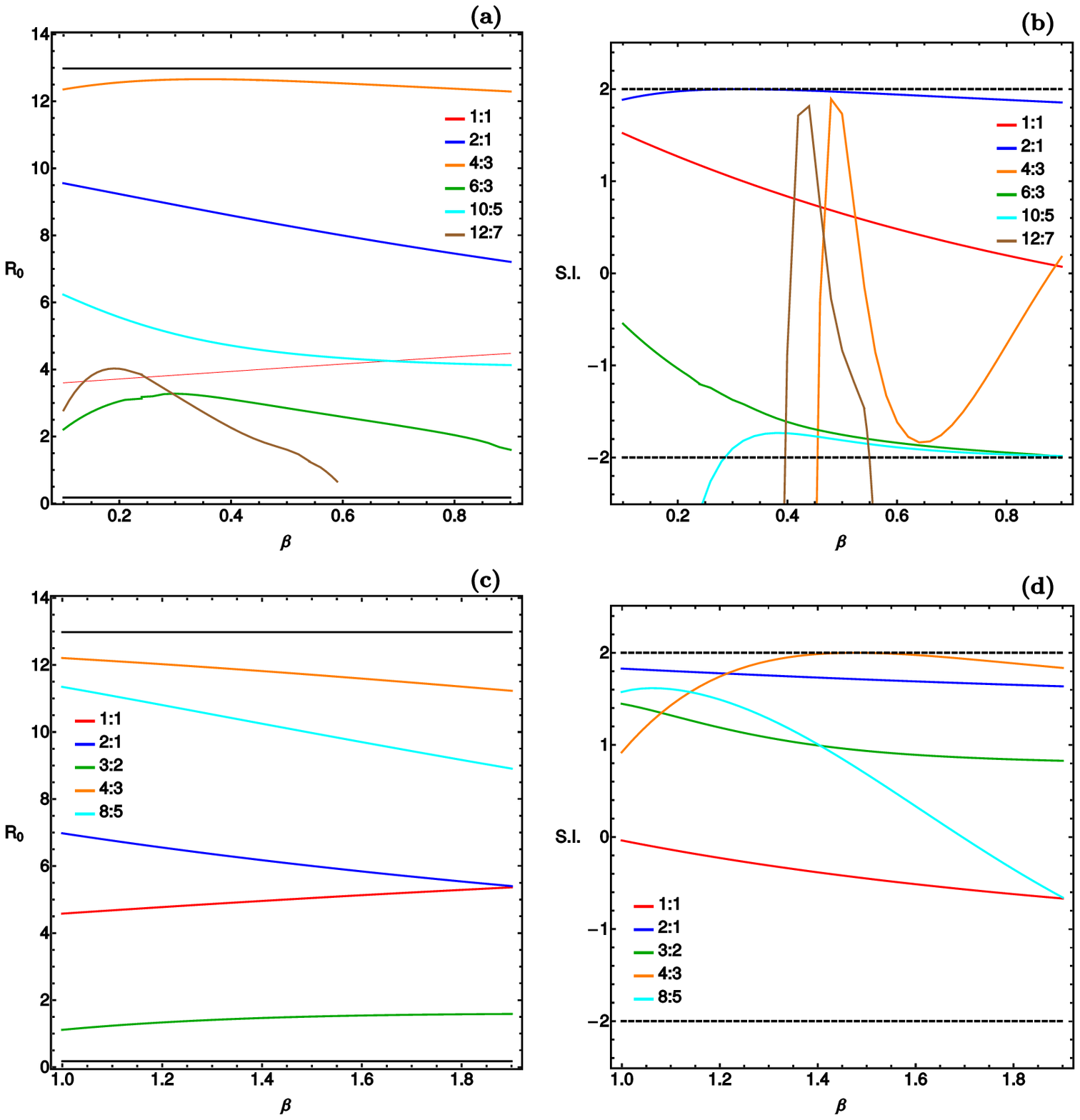}}}
\captionb{7}{Evolution of the $R$ coordinate of periodic orbits as a
function of the halo flattening parameter $\beta$ for the PH models
(panel a) and OH models (panel c), and the evolution of the
stability index S.I. as function of $\beta$ for the PH models (panel
b) and OH models (panel d).} \label{fig7}
\end{figure*}

\subsectionb{4.7}{Influence of the halo flattening parameter}

The exact shape (prolate, spherical, or oblate) of the dark matter halo
is determined by the flattening parameter $\beta$.  Thus, it would be of
particular interest to define how this parameter influences the position
and the stability of the periodic orbits in our PH and OH galaxy
models.  We let this parameter vary while fixing the values
of all the other parameters of our galactic models and integrating
orbits in the meridional plane for the set $\beta =
\{0.1,0.2,0.3,...,1.9\}$.

Fig.\,\ref{fig7}a shows the evolution of the positions of periodic
orbits as a function of the halo flattening parameter for the PH
models, while the stability of the same families in given in
Fig.\,\ref{fig7}b.  We observe that the 4:3 resonant family is
unstable for $0.1 \leq \beta \leq 0.446$, while the 10:5 secondary
resonant family becomes unstable in the interval $0.1 \leq \beta
\leq 0.281$.  Furthermore, the 12:7 resonant family is unstable for
$0.1 \leq \beta \leq 0.387$ and also for $0.538 \leq \beta \leq
0.592$, while it terminates at $\beta = 0.592$.  The evolution of
the $R$ coordinate of periodic orbits as a function of $\beta$ is
shown in panel (c).  The corresponding stability diagram in panel
(d) suggests that all resonant families are stable except for the
4:3 family which is almost critically stable for about $1.35 < \beta
< 1.7$.

\begin{figure*}[!th]
\centerline{ \resizebox{0.90\hsize}{!}{\includegraphics{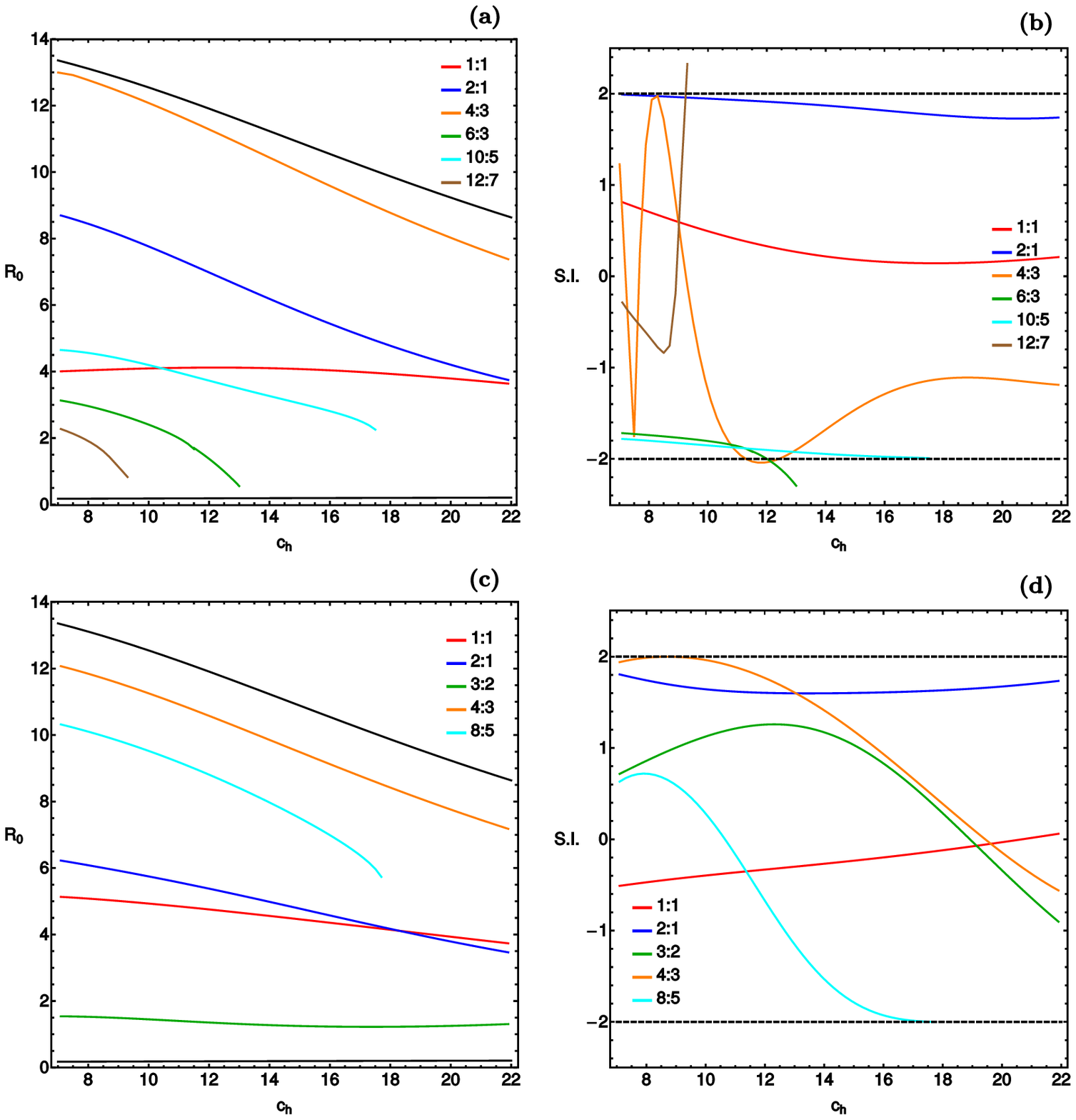}}}
\captionb{8}{Evolution of the $R$ coordinate of periodic orbits as a
function of the scale length of the halo, $c_{\rm h}$, for the PH
models (panel a) and OH models (panel c), and the evolution of the
stability index S.I. as a function of $c_{\rm h}$ for the PH models
(panel b) and OH models (panel d).} \label{fig8}
\end{figure*}

\subsectionb{4.8}{Influence of the scale length of the halo}

The concentration of the logarithmic, flattened dark matter halo is
controlled by its scale length $c_{\rm h}$.  In the following, we try to
reveal how the scale length of the dark halo influences the position and
the stability in our PH and OH galaxy models.  We let this
parameter vary while fixing the values of all other parameters of
our galactic models and integrating orbits in the meridional plane for
the set $c_{\rm h} = \{7,8.5,10,...,22\}$.

The evolution of the $R$ coordinate of periodic orbits as a function
of the scale length of the dark matter halo for the PH models is
given in Fig.\,\ref{fig8}a.  The corresponding stability diagram
(panel b) reveals that the 4:3 resonant family is unstable for
$11.29 \leq c_{\rm h} \leq 12.35$, while the 6:3 resonant family
becomes unstable for $12.07 \leq c_{\rm h} \leq 13.02$ and
terminates at $c_{\rm h} = 13.02$.  Moreover, the 10:5 resonant
family terminates at $c_{\rm h} = 17.51$ without first exhibiting
any signs of instability.  On the other hand, the 12:7 resonant
family becomes unstable in the interval $8.98 \leq c_{\rm h} \leq
9.25$ and terminates at $c_{\rm h} = 9.25$. Similarly, panel (c)
shows the evolution of the position of periodic orbits as a function
of $c_{\rm h}$ for the OH models.  The stability of the resonant
families is illustrated in panel (d), where we can see that all
families are stable, while the 4:3 resonant family is almost
critically stable for about $7 < c_{\rm h} < 9.5$. Furthermore, the
8:5 resonant family terminates at $c_{\rm h} = 17.72$ without
showing first any unstable branch.

\begin{figure*}[!th]
\centerline{ \resizebox{0.90\hsize}{!}{\includegraphics{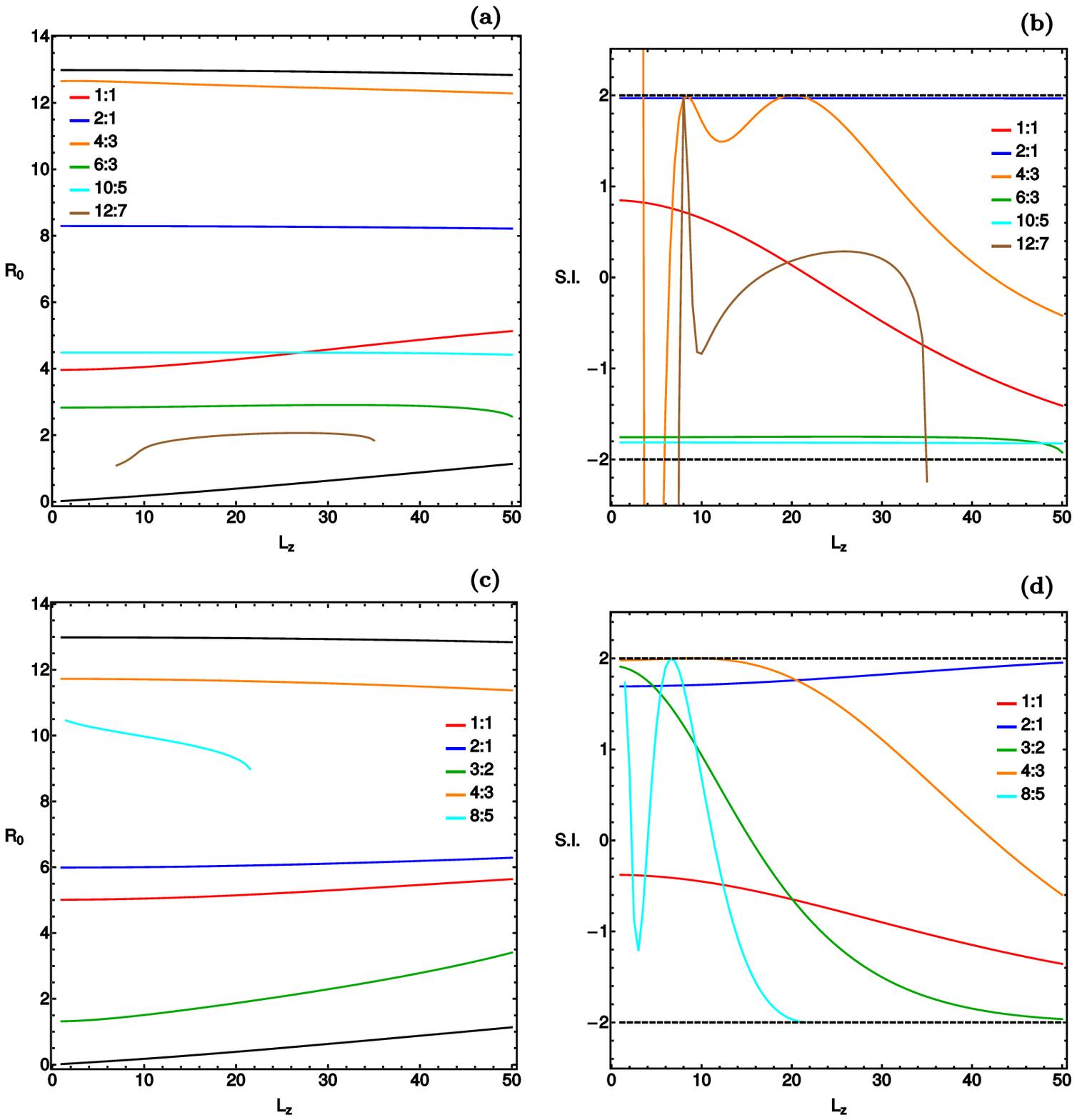}}}
\captionb{9}{Evolution of the $R$ coordinate of periodic orbits as a
function of the angular momentum $L_{\rm z}$ for the PH models
(panel a) and OH models (panel c), and the evolution of the
stability index S.I. as function of $L_{\rm z}$ for the PH models
(panel b) and OH models (panel d).} \label{fig9}
\end{figure*}

\subsectionb{4.9}{Influence of the angular momentum}

One of the most important quantities, which plays a vital role in the
nature of star orbits in the meridional plane $(R,z)$, is the angular
momentum $L_{\rm z}$.  Therefore, it is of paramount significance to
investigate how the angular momentum affects the position and the
stability of periodic orbits in our PH and OH galaxy models.  We let
this quantity vary while fixing the values of all the other parameters
of our galactic models and integrating orbits in the meridional plane
for the set $L_{\rm z} = \{1,5,10,...,50\}$.

In Fig.\,9a we present how the position of  periodic orbits in our PH
models evolves as a function of the angular momentum.  The
corresponding stability diagram in panel (b) shows that
the 4:3 resonant family becomes unstable in the interval $1.62 \leq
L_{\rm z} \leq 5.51$.  Furthermore, the 12:7 resonant family is unstable
for $6.34 \leq L_{\rm z} \leq 7.12$ and for $34.86 \leq L_{\rm z} \leq
35.09$, while the same family disappears near the two opposite ends of
the range of values of $L_{\rm z}$.  Panel (c) shows the
evolution of the $R$ coordinate of periodic orbits of the OH models
as a function of the angular momentum $L_{\rm z}$.  All resonant
families are present in the entire range of values except for
the 8:5 resonant family which terminates at $L_{\rm z} = 21.53$ without
first exhibiting any unstable branch.

\begin{figure*}[!th]
\centerline{
\resizebox{0.90\hsize}{!}{\includegraphics{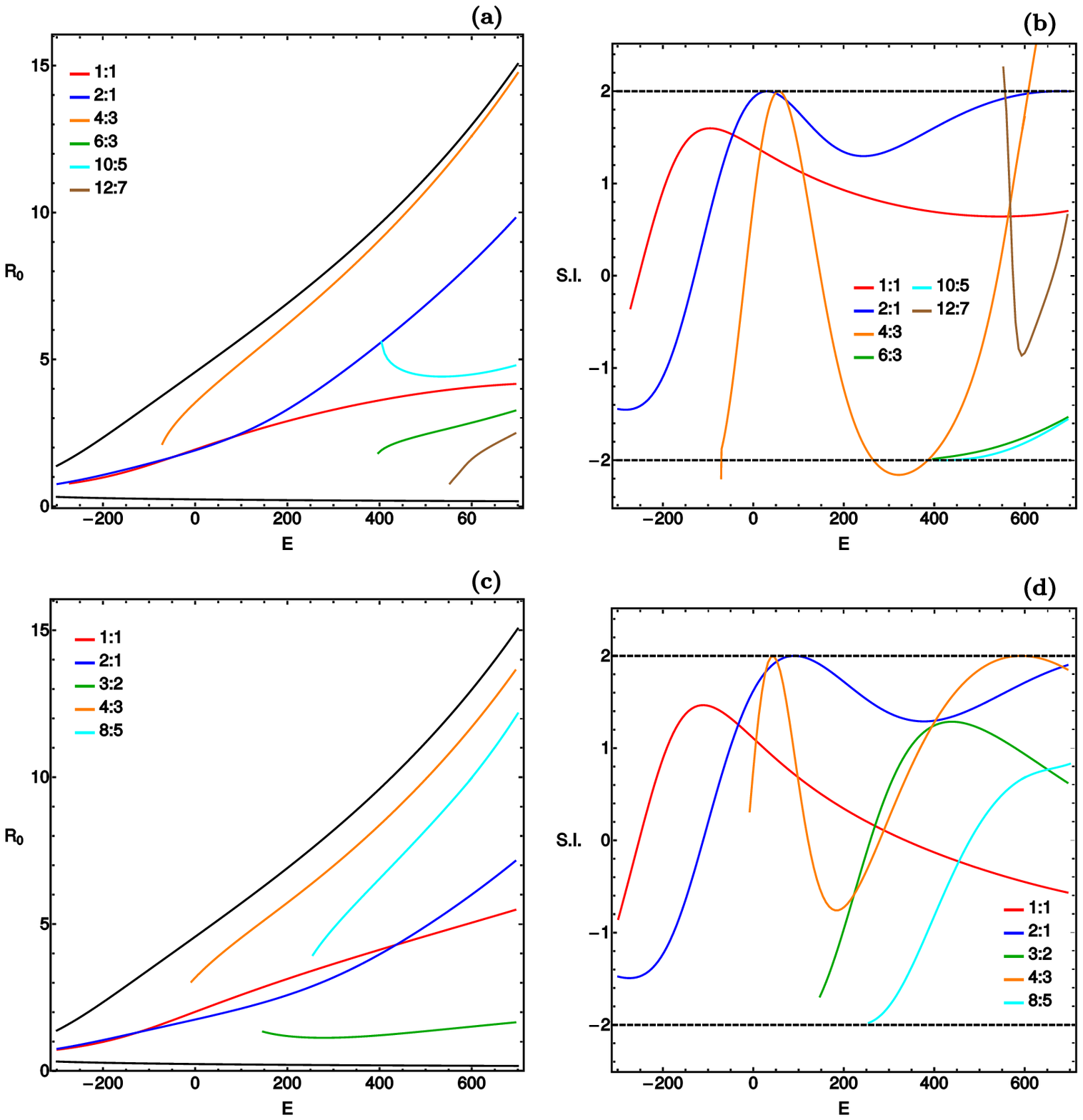}}}
\captionb{10}{Evolution of the $R$ coordinate of periodic orbits as
a function of the total orbital energy $(E)$ for the PH models
(panel a) and OH models (panel c), and the evolution of the
stability index S.I. as a function of $E$ for the PH models (panel
b) and OH models (panel d).} \label{fig10}
\end{figure*}

\subsectionb{4.10}{Influence of the orbital energy}

The last parameter under investigation is the total orbital energy $E$.
To explore how the energy level affects the position and the stability
of periodic orbits in our PH and OH galaxy models, we use the normal
procedure according to which we let the energy vary while fixing the
values of all the other parameters of our galactic models and
integrating orbits in the meridional plane for the set $E =
\{-300,-200,-100,0, 100,...,700\}$.  At this point we should point out
that the particular value of the energy determines the maximum possible
value of the $R$ coordinate $(R_{\rm max})$ on the $(R,\dot{R})$ phase
plane.  The energy values in the above interval result in $1.4 < R_{\rm
max} < 15$.

The evolution of the position of the periodic orbits as a function
of the total orbital energy in our PH models is presented in
Fig.\,\ref{fig10}a.  We should point out three aspects:  (i) for low
values of the orbital energy $(E < -100)$, which correspond to local
motion around the nucleus, only two main resonant families (2:1 and
1:1) survive; (ii) all the secondary resonant families (i.e, 6:3,
10:5 and 12:7) emerge at relatively high energy levels $(E
> 400)$ which correspond to global motion; (iii) at about $E = 400$
we observe how the 10:5 secondary resonant family bifurcates from
the main 2:1 family.  In panel (b) of the same figure we present the
corresponding stability diagram from which we can deduce the
following: (i) the 4:3 resonant family is unstable in the intervals
$-71.32 \leq E \leq -70.98$, $265.75 \leq E \leq 385.06$ and $609.12
\leq E \leq 700$; (ii) the 6:3 and 10:5 resonant families terminate
at $E = 397.28$ and $E = 404.05$, respectively, without first
exhibiting any instability; (iii) the 12:7 resonant family starts at
$E = 552.71$ and remains unstable for $552.71 \leq E \leq 553.78$.
Panel (c) depicts the evolution of the $R$ coordinate of the
periodic orbits as a function of $E$ for the OH models.  The
structure of the characteristic diagram is quite similar to that
discussed earlier in panel (a).  The stability diagram given in
panel (d) shows that all resonant families are stable. In addition
one can see that the 3:2, 4:3 and 8:5 resonant families originate at
$E = 147.28$, $E = -8.10$ and $E = 255.41$, respectively.

\sectionb{5}{DISCUSSION}
\label{disc}
\vskip-2mm

We used the composite analytic axially symmetric galactic gravitational
model introduced in Paper I in order to reveal the complete network of
periodic orbits.  Varying the values of all the involved parameters of
the dynamical system, as well as the two global isolating integrals of
motion, namely the angular momentum and the total orbital energy, we
managed to construct detailed and aggregated diagrams showing the
evolution of the position ($R$ coordinate) and the stability of the main
resonant periodic families of orbits as a function of the variable
parameters.  Here we should emphasize that this is the first such
detailed and systematic numerical investigation regarding the influence
of all the dynamical quantities of the system on the position and
stability of resonant periodic families.

In order to locate the position and measure the stability of the
periodic orbits, we used a highly sensitive numerical routine which has
the ability to identify resonant periodic orbits of the type $n:m$,
where $m$ and $n$ are the main frequencies of a periodic orbit along the
$R$ and $z$ axis, respectively.  First, for every variable parameter a
range of realistic values was defined.  The numerical code begins from
the value corresponding to the standard model (SPM and SOM) which is
always somewhere inside the main interval and then using a variable step
scans all the available interval thus calculating the entire resonant
family.  Two different cases were investigated for every parameter:  (i)
the case where the dark matter halo component is prolate and (ii) the
case where an oblate dark matter halo is present.

Our numerical computations revealed that all the dynamical
quantities affect, more or less, the position as well as the
stability of the resonant periodic orbits.  It was observed,
however, that for both types of models (PH and OH) the mass of the
nucleus, the mass of the disk, the halo flattening parameter, the
scale length of the halo, the angular momentum, and the total
orbital energy are the most influenced quantities, while the effects
of the scale length of the nucleus and that of the horizontal and
vertical scale length of the disk are much weaker.

\thanks{ I would like to express my warmest thanks to Dr.  Leonid P.
Ossipkov (University of Saint Petersburg) for the careful reading
the manuscript and for all the apt suggestions and comments that
allowed me to improve both the quality and the clarity of the
paper.}

%\newpage

%\renewcommand*{\refname}{}
\References
%\vspace*{2\baselineskip}

\begingroup
\renewcommand{\section}[2]{}

\endgroup


\begin{thebibliography}{}

\parskip=-7pt

\bibitem[\protect\citeauthoryear{Abalakin}{1961}]{A61} Abalakin V. K.
1961, Bull.  Inst. Theor.  Astron., 8, 173

\bibitem[\protect\citeauthoryear{Abalakin}{1963}]{A63} Abalakin V. K.
1963, Bull. Inst. Theor. Astron., 9, 204

\bibitem[\protect\citeauthoryear{Antonov}{1974}]{A74} Antonov V. A.
1974, Trans. Astron. Obs. Lenigrad State Univ., 30, 111

\bibitem[\protect\citeauthoryear{Allen \& Santill\'an}{1991}]{AS91}
Allen C., Santill\'an A. 1991, Rev.  Mex. Astron.  Astrof., 22, 255

\bibitem[\protect\citeauthoryear{Belbruno et al.}{1994}]{BLO94}
Belbruno E., Llibre J., Oll\'{e} M. 1994, Celest.  Mech., 60, 99

\bibitem[\protect\citeauthoryear{Binney \& Tremaine}{2008}]{BT08}
Binney J., Tremaine S. 2008, {\it Galactic Dynamics}, Princeton
Univ. Press

\bibitem[\protect\citeauthoryear{Bollt et al.}{1997}]{BLG97} Bollt
E. M., Lai Y. C., Grebogi  C. 1997, Phys.  Rev.  Lett, 63, 3787

\bibitem[\protect\citeauthoryear{Carlberg \& Innanen}{1987}]{CI87}
Carlberg R. G., Innanen K. A. 1987, AJ, 94, 666

\bibitem[\protect\citeauthoryear{Carpintero \& Aguilar}{1998}]{CA98}
Carpintero D. D., Aguilar L. A. 1998, MNRAS, 298, 1

\bibitem[\protect\citeauthoryear{Contopoulos \& Barbanis}{1985}]{CB85}
Contopoulos G., Barbanis B. 1985, A\&A, 153, 44

%\bibitem[\protect\citeauthoryear{Contopoulos \& Barbanis}{1985}]{CB94}
%Contopoulos G., Barbanis B. 1994,  Celest. Mech., 59, 279

\bibitem[\protect\citeauthoryear{Contopoulos \& Magnenat}{1985}]{CM85}
Contopoulos G., Magnenat P. 1985, Celest. Mech., 37, 387

\bibitem[\protect\citeauthoryear{Contopoulos \& Mertzanides}{1977}]{CM77}
Contopoulos G., Mertzanides C. 1977, A\&A, 61, 477

\bibitem[\protect\citeauthoryear{Contopoulos \& Seimenis}{1990}]{CS90}
Contopoulos G., Seimenis J. 1990, A\&A, 227, 4

\bibitem[\protect\citeauthoryear{Cvitanovi\'{c}}{1991}]{C91}
Cvitanovi\'{c} P. 1991, Physica D, 51, 138

\bibitem[\protect\citeauthoryear{Davidchack et al.}{2000}]{DLBD00}
Davidchack R. L., Lai Y. C., Bollt E. M., Dhamala M. 2000, Phys.  Rev.
E, 61, 1353

\bibitem[\protect\citeauthoryear{Devaney}{1989}]{D89} Devaney R. L.
1989, {\it An Introduction to Chaotic Dynamical Systems}, 2nd ed.,
Addison-Wesley Publishing Co.

\bibitem[\protect\citeauthoryear{Deprit \& Henrard}{1967}]{DH67} Deprit
A., Henrard J. 1967, AJ, 72, 158

\bibitem[\protect\citeauthoryear{G\'{o}mez et al.}{1985}]{GLMS85}
G\'{o}mez G., Llibre J., Mart\'{i}nez R., Sim\'{o} C. 1985, {\it Station
Keeping of Libration Points}, ESOC Conf. 5648/83/D/JS(SC)

\bibitem[\protect\citeauthoryear{Gutzwiller}{1990}]{G90} Gutzwiller
M. C. 1990, {\it Chaos in Classical and Quantum Mechanics}, Springer
Verlag

\bibitem[\protect\citeauthoryear{Henrard \& Lemaitre}{1986}]{HL86}
Henrard J., Lemaitre A. 1986, Celest.  Mech., 39, 213

\bibitem[\protect\citeauthoryear{Howell}{1984}]{H84} Howell K. C. 1984,
Celest.  Mech., 32, 53

\bibitem[\protect\citeauthoryear{Karimov \& Sokolsky}{1989}]{KS89}
Karimov S. R., Sokolsky A. G. 1989, Celest.  Mech., 46, 335

\bibitem[\protect\citeauthoryear{Lara \& Pal\'{a}ez}{2002}]{LP02} Lara
M., Pal\'{a}ez J. 2002, A\&A, 389, 692

\bibitem[\protect\citeauthoryear{Lathrop \& Kostelich}{1989}]{LK89}
Lathrop D. P., Kostelich E. J. 1989, Phys.  Rev. A, 40, 4028

\bibitem[\protect\citeauthoryear{Miyamoto \& Nagai}{1975}]{MN75}
Miyamoto W., Nagai R. 1975, PASJ, 27, 533

\bibitem[\protect\citeauthoryear{Ott}{1993}]{O93} Ott E. 1993, {\it
Chaos in Dynamical Systems}, Cambridge University Press

\bibitem[\protect\citeauthoryear{Ott et al.}{1990}]{OGY90} Ott E.,
Grebogi C., Yorke J. A. 1990, Phys.  Rev.  Lett, 64, 1196

\bibitem[\protect\citeauthoryear{Plumecoq \& Lefranc}{2000}]{PL00}
Plumecoq J., Lefranc M. 2000, Physica D, 144, 231

\bibitem[\protect\citeauthoryear{Prendergast}{1982}]{P82} Prendergast K.
H. 1982, in {\it The Riemann Problem}, Lecture Notes in Mathematics,
Springer-Verlag, 925, 369

\bibitem[\protect\citeauthoryear{Press et al.}{1992}]{PTVF92} Press H.
P., Teukolsky S. A., Vetterling W. T., Flannery B. P. 1992, {\it
Numerical Recipes in FORTRAN 77}, 2nd ed., Cambridge Univ.  Press

\bibitem[\protect\citeauthoryear{Scheeres}{1999}]{S99} Scheeres D. J.,
1999, in {\it Hamiltonian Systems with Three or More Degrees of
Freedom}, NATO AS I Ser., vol. 533, p. 554

\bibitem[\protect\citeauthoryear{Wood}{1984}]{W84} Wood D. 1984, IMA
J. Appl.  Math., 33, 229

\bibitem[\protect\citeauthoryear{Zotos}{2013}]{Z13} Zotos E. E. 2013,
Nonlinear Dynamics, 73, 931

\bibitem[\protect\citeauthoryear{Zotos}{2014}]{Z14} Zotos E. E. 2014,
A\&A, 563, A19 (Paper I)

\bibitem[\protect\citeauthoryear{Zotos \& Carpintero}{2013}]{ZC13} Zotos
E. E., Carpintero D. D. 2013, CeMDA, 116, 417

\end{thebibliography}
\end{document}